# Stochastic accretion of the Earth

Paolo A. Sossi[1*], Ingo L. Stotz[2], Seth A. Jacobson[3], Alessandro Morbidelli[4], Hugh St.C. O'Neill[5]

[1] *Institute of Geochemistry and Petrology, ETH Zürich, CH-8092, Zürich, Switzerland*

[2] *Institut für Geophysik, Ludwig-Maximilians Universität München, München, Germany*

[3] *Department of Earth and Environmental Sciences, Michigan State University, East Lansing, MI 48824, USA*

[4] *Laboratoire Lagrange, UMR7293, Université de Nice Sophia-Antipolis, CNRS, Observatoire de la Côte d'Azur, Boulevard de l'Observatoire, 06304 Nice Cedex 4, France*

[5] *School of Earth, Atmosphere and Environment, Monash University, Melbourne, VIC 3800, Australia*

*\* correspondence to: [paolo.sossi@erdw.ethz.ch](mailto:paolo.sossi@erdw.ethz.ch)*

**Earth is depleted in volatile elements relative to chondritic meteorites, its possible building blocks. The extent of this depletion increases with decreasing condensation temperature, and is approximated by a cumulative normal distribution, unlike that in any chondrite. However, moderately volatile elements, occupying the mid-range of the distribution, have chondritic isotope ratios, contrary to that expected from loss by partial vaporisation/condensation. Here we reconcile these observations by showing, using N-body simulations, that Earth accreted stochastically from many precursor bodies whose variable compositions reflect the temperatures at which they formed. Impact-induced atmospheric loss was efficient only when the proto-Earth was small, and elements that accreted thereafter retain near-chondritic isotope ratios. Earth's composition is reproduced when initial temperatures of planetesimal- to embryo-sized bodies are set by disk accretion rates of $(1.08\pm0.17)\times10^{-7}$ solar masses/yr, although they may be perturbed by $^{26}$Al heating on bodies formed at different times. The model implies a heliocentric gradient in composition and rapid planetesimal formation within ~1 Myr, in accord with radiometric volatile depletion ages of Earth.**

Terrestrial planets of our Solar System, including Earth, are thought to represent the sum of progressive collisional growth of smaller bodies, whose size distribution evolved through time and space. In the classical model, collisions involve mostly nearby objects and lead to the formation of tens of Moon- to Mars-sized embryos within the lifetime of the protoplanetary disk (a few Myr)[1]. Only after the disappearance of gas from the disk, which causes embryos to become dynamically unstable, is mixing between objects from different heliocentric distances expected over $10^7$ to $10^8$ yr timescales[2,3]. Alternatively, recent models suggest that the accretion of most of the Earth's mass occurred within the lifetime of the disk via the capture of sunward-drifting cm-sized pebbles that initially formed at the 'snow-line'[4,5].

Such models of Earth's accretion may be tested by examining the chemical and isotopic composition of the Bulk Silicate Earth (BSE), which has been estimated from geological evidence[6]. The BSE is the Earth excluding its metallic core, which is inaccessible to detailed chemical scrutiny. In the BSE, refractory lithophile elements (RLEs[7]) occur in solar proportions relative to one another, whereas the abundances of moderately volatile lithophile elements decrease according to their calculated 50 % nebular condensation temperatures ($T_c^{50}$) (Fig. 1). These elements have $T_c^{50} < 1300$ K and include the lithophile elements K, Rb, Zn and Cl, whose isotopic constitutions are similar to those measured in chondrites[8–11]. This observation precludes their depletion as being the result of partial evaporative loss from the Earth or its precursors, which would have caused detectable isotopic fractionation[12,13].

Instead of partial evaporation, element depletion would occur without isotopic fractionation if the Earth accreted from mixtures of components in which an element is either present in solar abundances relative to RLEs (and therefore with chondritic isotope ratios) or almost entirely absent, with the likelihood of its presence/absence being a function of its condensation temperature. As most elements are calculated to condense from the solar nebula gas over a narrow temperature window (a few tens of Kelvin), the condensation of one element should be nearly complete before that of the next element has begun[13,14]. If condensation ceases at some threshold temperature, $T_0$, a step-function pattern in the planetary body is produced when the CI- (Ivuna-type carbonaceous chondrites), Al-normalised abundance of element $i$, $f_i$, is plotted against $T_c^{50}$ [15]. Because such step-like depletion patterns are observed in small, rocky bodies (*e.g.,* Vesta; Fig. 1), this feature was likely commonplace in the pebble- and planetesimal-sized building blocks of Earth. The Earth instead exhibits a smoother decline in elemental abundance over a wider range of $T_c^{50}$ (Fig. 1) that differs from any chondrite. Indeed, no mixture of the extant suite of chondritic meteorites can simultaneously reproduce the isotopic and chemical composition of the Earth[16,17].

Elemental abundances ($x$) normalised to CI and Al in both the Earth and Vesta increase with increasing $T_c^{50}$ in a manner approximated by a cumulative normal distribution;

$$f_i^{CI,Al} = \frac{(x_i/x_{Al})_{body}}{(x_i/x_{Al})_{CI}} = \frac{1}{2}\left[1 + erf\left(\frac{T_c^{50} - T_0}{\sigma\sqrt{2}}\right)\right] (1)$$

The inflection point reflects the mean temperature, $T_0$, of the distribution. Weighted fits yield 1144±12 K and 1081±29 K, for Earth and Vesta, respectively. The steepness of the sigmoid is described by $\sigma$, the standard deviation, approximating a step-function at low $\sigma$ (57±17 K, Vesta) or a gradual curve at high $\sigma$ (226±16 K, Earth).

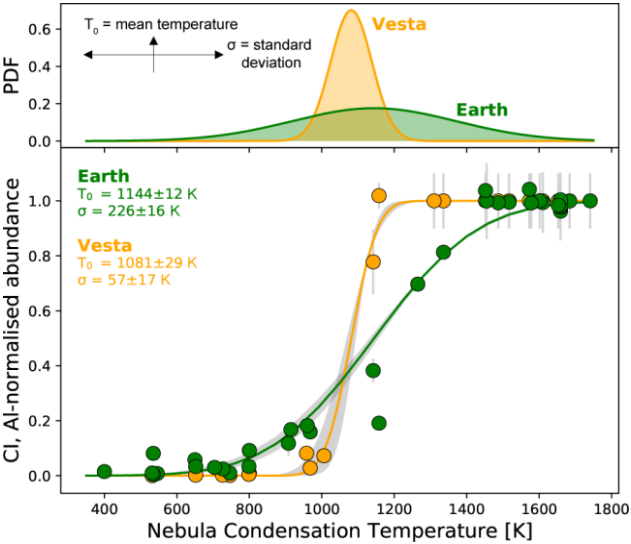

**Figure 1 | Fits to the bulk compositions of the mantles of Earth and Vesta.** CI-, Al-normalised abundances ($f_i$) and $1\sigma$ uncertainties of lithophile elements in Bulk Silicate Earth[6] (green) and Vesta[18] (yellow) plotted as a function of their 50 % nebular condensation temperatures[19], with Si, Na, K, Zn and Cl updated from ref. [20]. Both bodies have abundances that approximate a cumulative normal distribution (eq. 1). The corresponding probability distribution function (PDF) is shown in the top panel. The curves are determined by a least-squares fit to error-weighted elemental abundances, $\chi^2 = \left([f_i - \bar{f}_i]/s_i\right)^2$ where $\bar{f}_i$ is the modelled value, $f_i$ is the abundance, and $s_i$ its $1\sigma$ uncertainty (Extended Data Table 2). Best-fit values of $T_0$ and $\sigma$ are listed on the plot and yield Root-Mean Squared deviations of 0.071 ($n = 35$) for Earth and 0.028 ($n = 25$) for Vesta. Manganese falls significantly below the trend for Earth's mantle, likely owing to partial incorporation into Earth's core[21]. The linear scale highlights gradual elemental depletion with $T_c^{50}$ in the BSE, arguing against scenarios in which the Earth formed by mixing of two components (85% volatile-poor; 15 % volatile-rich)[22,23]. A log-scale version of Fig. 1 is shown in Extended Data Fig. 3.

The mean temperatures experienced by the building blocks of Earth (1 AU) exceeded those of Vesta (2.36 AU), consistent with, but not indicative of, a heliocentric gradient in composition. Any such gradient among precursor bodies may have been subsequently perturbed by the $^{26}$Al heating they experienced and/or their specific orbital and collisional histories. The higher standard deviation in $T_0$ of the Earth indicates that it collected material with a wider range of temperatures than did Vesta. This result is in accord with dynamical models that predict larger bodies accumulate material from a broader range of heliocentric distances than do smaller ones[2,3]. On this basis, we propose that Earth did not accrete from chondrites, but rather stochastically from a summation of smaller precursor bodies, each with Vesta-like step-function element abundance patterns ($\sigma$ tends to 0) and variable threshold temperatures ($T_0$) that reflect the temperatures at which they formed.

In order to quantitatively test this hypothesis, we examine four different N-body simulations[24] of the 'Grand Tack' scenario that reproduce the approximate number, masses and semi-major axes of the terrestrial planets. In this class of models, the migration of Jupiter, first inwards, then outwards by its resonant interaction with Saturn clears the region between 3 to 6 Astronomical Units (AU) of planetesimals and embryos, and scatters them into the inner disk between 0.7 and 3 AU, leaving a remnant population of planetesimals in the asteroid belt and beyond 6 AU[24,25] (Table 1). Any other dynamical model resulting in a comparable radial distribution of accreted material (Extended Data Fig. 4) would give equivalent results.

The sensitivity of the chemical and isotopic make-up of the Earth to the compositional model, described below, was evaluated by performing 1000 Monte-Carlo (MC) simulations with three free parameters for each N-body run (see *Methods*). It is assumed that all precursor bodies formed at the same time, meaning their initial temperature ($T_0$) reflects their heliocentric distance, $d$, and is calculated according to the radial temperature profile of a high-opacity nebular midplane, $T_{midplane}$, set by the disk in steady-state mass accretion[26,27] (Extended Data Fig. 5a):

$$T_0 = T_{midplane} = \left(\frac{9\tau G M_\odot \dot{M}}{64\pi\sigma_B d^3}\right)^{\frac{1}{4}}, (2)$$

where $\tau$ is the optical depth at the midplane (see *Methods*), $M_\odot$ is the mass of the Sun and $\sigma_B$ the Stefan-Boltzmann constant. The mass accretion rate to the Sun, ($\dot{M}$), the first free parameter, is varied randomly within a normal distribution of 1(±0.25, 1$\sigma$)×10$^{-7}$ $M_\odot$/yr. Observations of T-Tauri stars show that $\dot{M}$ ($M_\odot/yr$) decays as a function of time (Myr)[28]:

$$\log(\dot{M}) = -8.00 \pm 0.10 - (1.40 \pm 0.29)\log(t), (3)$$

so that its value indicates the time of planetesimal- or embryo formation. Given $T_{midplane}$, planetesimals contain CI-normalised abundances ($f_i$) of 30 fictive elements, defined with $T_c^{50}$ (at $10^{-4}$ bar) ranging from 350 to 1800 K at 50 K intervals (see *Methods*, Extended Data Table 1):

$$f_i = x_{i,body}/x_{i,CI}, (4)$$

| Simulation | Embryos (0.7 - 3 AU) | | Planetesimals (0.7 – 3 AU) | | Planetesimals (6 – 8 AU) | | Earth | |
|---|---|---|---|---|---|---|---|---|
| | Mass ($M_E$) | $n$ | Mass ($M_E$) | $n$ | Mass ($M_E$) | $n$ | Mass ($M_E$) | Semi-Major Axis (AU) |
| GT/8:1/0.025/R2 | 0.025 | 212 | $3.81\times10^{-4}$ | 1750 | $3.81\times10^{-4}$ | 500 | 1.107 | 0.936 |
| GT/4:1/0.025/R7 | 0.025 | 170 | $3.81\times10^{-4}$ | 2783 | $4.67\times10^{-5}$ | 1563 | 0.957 | 0.907 |
| GT/4:1/0.05/R8 | 0.05 | 87 | $3.81\times10^{-4}$ | 2836 | $5.15\times10^{-5}$ | 1563 | 0.937 | 0.969 |
| GT/8:1/0.08/R18 | 0.08 | 67 | $3.81\times10^{-4}$ | 1750 | $3.45\times10^{-4}$ | 500 | 1.074 | 1.179 |

Table 1 | Input parameters and final masses and semi-major axes of the analogue planet of the Earth resulting from the N-body simulations. The naming convention of N-body simulations follows *i)* the Grand Tack (GT) model, *ii)* the mass ratio of embryos:planetesimals, *iii)* the mass of embryos in $M_E$ and *iv)* the run number with those parameters[25].

where $x$ is the mole fraction. Fictive elements render the models independent of uncertainties or revisions in $T_c^{50}$ of real elements[19,29], and permit quantification of systematic trends in their behaviour. If $T_{midplane} > T_c^{50}$ of element $i$, $f_i = 0$, otherwise, $f_i = 1$, defining a 'step-function' abundance pattern in planetesimals. This distribution reflects *i)* their thermodynamic properties resulting in their perfect retention or evaporation[15] and *ii)* the inability of small (< 1000 km) bodies to retain an atmosphere[30]. The embryos comprise numerous planetesimals, each with a $T_{midplane}$ sampled normally about the $T_{midplane}$ of the embryo with a standard deviation, $\sigma_{embryo}$ (*cf.* eq. 1). This, the second free parameter, has a mean of 87 K with standard deviation of 24 K (1σ) to determine the sensitivity to the smoothness of the embryo element patterns. This temperature range equates to a feeding zone width[31] of roughly ±0.1 AU for an embryo forming at 1 AU.

To assess the extent to which partial evaporation, and hence isotopic fractionation occurs during planetary accretion, the model computes a surface temperature increase in each body experiencing an impact that is proportional to the kinetic energy imparted by that impact (see *Methods*). Each fictive element vaporises with a partial pressure determined by its thermodynamic properties, with more volatile elements evolving higher partial pressures (see *Methods*). Given the surface atmospheric temperature and pressure, the temperature-pressure profile of a convective atmosphere with height following the dry adiabat is determined, before transitioning to an isothermal region above the homopause[32]. Elements in the atmosphere may be lost to space *via* Jeans and hydrodynamic escape. Which regime prevails depends on the escape parameter (Extended Data Fig. 6):

$$\lambda_{esc} = \frac{v_{esc}^2 m_i}{2k_B T_{esc}}, (5)$$

where $v_{esc}$ is the escape velocity of the body (= $\sqrt{2GM_{body}/r_{body}}$), $k_B$ is the Boltzmann constant, $T_{esc}$ is the temperature at the escaping surface and $m_i$ is the mean atmospheric mass. The value of $m_i$, the third free parameter, is sampled randomly from a normal distribution with mean 33.5 g/mol and standard deviation = 3.5 g/mol, with limits corresponding to 23 g/mol (Na) and 44 g/mol (SiO). Jeans escape occurs for $\lambda_{esc} > 3$, whereas hydrodynamic escape prevails for $\lambda_{esc} < 3$ (see *Methods*)[33], while escape by other, non-thermal processes such as photoevaporation or ionisation are neglected. Escape fluxes are calculated at 10 K steps, assuming grey-body cooling of the atmosphere (see *Methods*). The composition of the residue is re-computed at each cooling step, and assumes element abundances equilibrate in the entire body.

In all simulations, refractory elements accrete in concert with the total mass of Earth (Fig. 2; Extended Data Figs. 7, 8), because they are present in all planetesimals and are unaffected by atmospheric losses. By contrast, more volatile elements lag behind their refractory counterparts. Simulations with high atmospheric loss rates (low $m_i$) are indistinguishable from cases in which atmospheric escape is suppressed (high $m_i$) (Extended Figure 11b), meaning the differential timing of volatile accretion depends only on precursor body compositions set by the value of $\dot{M}$ and its original heliocentric distance, $d$.

Our results indicate that, while the proto-Earth accreted ~80-90% of its final total mass by 25 – 100 Myr, only ~50 % of the most volatile elements ($T_c^{50} < 500$ K) accreted within this timeframe. This reflects the preferential accretion of volatile-undepleted (*i.e.,* CI-like) planetesimals from beyond 6 AU later in Earth's growth. Late accretion of this material (*e.g.,* during a giant impact) provides a physical explanation for the requirement that volatile-rich material constituted the final ~10 % of Earth's mass and accreted after 30 Myr in order to explain the Earth's carbonaceous chondrite-like $^{107}Ag/^{109}Ag$ ratio, and to reconcile Hf-W and Pd-Ag ages[34].

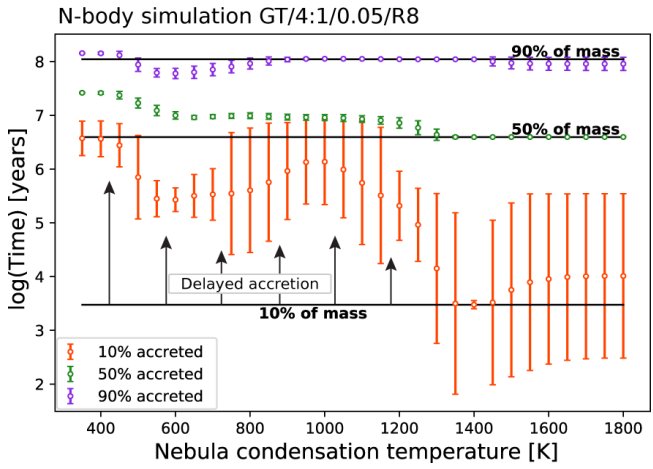

**Figure 2 | Timing of element accretion to the Earth as a function of volatility.** Points represent the time in the single N-body simulation (ordinate) at which a given element (abscissa) reaches a threshold value (10 %, red; 50 %, green; or 90 %, purple) of its final abundance. These points are % accreted = $100 \times (x_i M)_t/(x_i M)_f$, where the subscripts $t$ and $f$ denote the threshold time and the final time, respectively, $x$ is the mole fraction of element $i$ and $M$ is the mass of the growing Earth. Error bars (1 standard deviation, $1\sigma$) denote the range given by Monte Carlo models for different $\dot{M}$, $\sigma_{embryo}$ and $m_i$ across 1000 simulations. No error bar indicates that the element reaches the given % accreted value at the same time in each MC simulation. The mass of the Earth, $M_t/M_f$, through time is delineated by horizontal lines at 10 %, 50 % and 90 %. Offset between these lines and the coloured points is the differential accretion time. The scatter in refractory elements arises from their dilution/enrichment in planetesimals and embryos with/without the more abundant elements with $T_c^{50}$ near 1350 K (Extended Data Table 1).

Once fully accreted, modelled elemental abundances in the BSE adhere to a gradual decline with decreasing $T_c^{50}$ (Fig. 3a, b) that parallel observations, independent of the specific N-body simulation (Extended Data Fig. 9a-d). Although a constant-abundance plateau is observed in some simulations, it extends only to elements with $T_c^{50} < 450$ K, and never up to 750 K or 1000 K as suggested in recent models[17,23,35]. Monte-Carlo simulations indicate that the composition of the BSE depends strongly on $\dot{M}$ but is nearly independent of $\sigma_{embryo}$ and $m_i$ (Extended Data Figs. 11, 12). Therefore, its final composition is inherited from its precursors, rather than through subsequent modification by partial evaporation during collisions. This result follows from the condition applied to planetesimals that $f_i = 0$ when $T_{midplane} > T_c^{50}$, which becomes increasingly probable as $T_c^{50}$ decreases.

The elongated (high-$\sigma$) shape in the BSE arises from the summation of precursor bodies with a wide range of $T_{midplane}$, which, for a given $\dot{M}$, relates to heliocentric distance (Extended Data Fig. 5). Even though planetesimals initially have step-function patterns, this process produces a smooth pattern in the BSE by the Central Limit Theorem. The corollary is that step-function patterns (low-$\sigma$) should be more evident in smaller bodies that form in local feeding zones, in which precursors have a restricted range of $T_{midplane}$, as observed for Vesta[15] (Fig. 1).

Best fits to the lithophile elemental abundances in the BSE are determined by minima of the Root-Mean-Squared (RMS) deviation of models from the data (Fig. 3; Extended Data Fig. 9a-d). The RMS is minimised for $\dot{M} = 1.08(\pm 0.17, 1\sigma) \times 10^{-7}$ $M_\odot$/yr (Extended Data Fig. 12), and is insensitive to the specific accretion history in the subset of the four N-body simulations investigated here. Assuming gradual, rather than step-like, patterns for planetesimals (as per those of embryos) does not change the best fit value of $\dot{M}$. These mass accretion rates yield $T_{midplane} = 1410 \pm 60$ K at 0.7 AU, coinciding with the $T_c^{50}$ of the major rock-forming elements, Mg, Si and Fe[19].

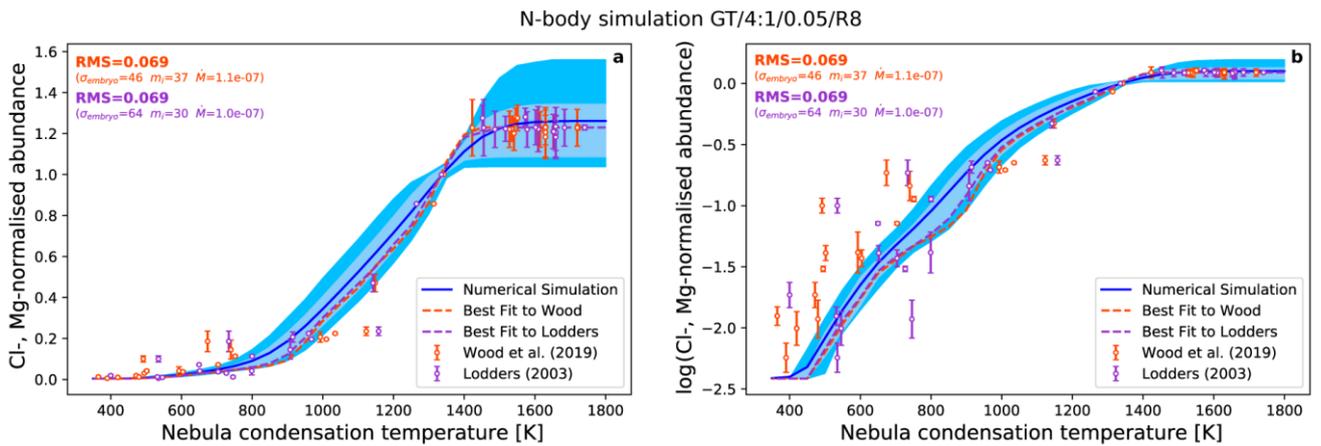

**Figure 3 | Elemental abundances in the fully accreted Earth.** CI-, Mg-normalised elemental abundances in the BSE[6] as a function of 50% nebula condensation temperatures according to Wood et al., red[29]; and Lodders with updates from Fegley and Schaefer, purple[19,20], compared with abundances predicted by 1000 Monte Carlo simulations per N-body simulation plotted on **a)** a linear scale and **b)** a logarithmic scale. The average composition over 1000 runs is given by the solid blue line with the percentile ranges of the simulations (25th to 75th, light blue and 10th to 90th, dark blue). The dashed lines show the simulations that minimise the Root-Mean-Square (RMS) deviation from the data. The RMS values are listed in the top left-hand corner, along with those of the three random variables to which they correspond. Only lithophile elements were plotted, as siderophile elements are further depleted by core formation, which is not considered in our model.

Observations of T-Tauri stars show that younger disks have faster mass accretion rates than do older ones[36] (eq. 3). By analogy, accretion rates of ~$10^{-7}$ $M_\odot$/yr indicate rapid establishment, ~0.2±0.1 Myr, of the compositional architecture of the inner disk; within the lifetime of the nebular gas. Detailed numerical models also support early formation of planetesimals in the inner disk at ~0.5 Myr[37] or ~1 Myr[38]. Moreover, this timescale matches volatile depletion ages based on Rb-Sr and Mn-Cr systematics of small differentiated bodies (<1.5 Myr[39,40]) and of the Earth (1-2 Myr[41]). Subsequent, post-nebular orbital evolution produces its smooth (high-σ) pattern over longer timescales, nearing ~$10^8$ yr.

The stable isotope composition of the Earth represents the sum of contributions from partial evaporative losses and accretion of unevaporated materials. The isotope fractionation evolved is proportional to the amount of the element lost and the fractionation factor. In the N-body simulations, isotope ratios for each element in planetesimals and embryos are initially CI chondritic, and atmospheric escape is calculated for all bodies that experience a collision. The magnitude of isotope fractionation in the fully accreted Earth is proportional to $1/m_i$ (Fig. 4, Extended Data Fig. 13). A peak is observed in elements with $450 < T_c^{50}$ (K) $< 800$ across all simulations (Fig. 4; Extended Data Fig. 10). These elements are not so volatile so as to have been lost completely (and thus prone to overprinting by late-accreted material) but still sufficiently volatile to record partial evaporation over Earth's accretion.

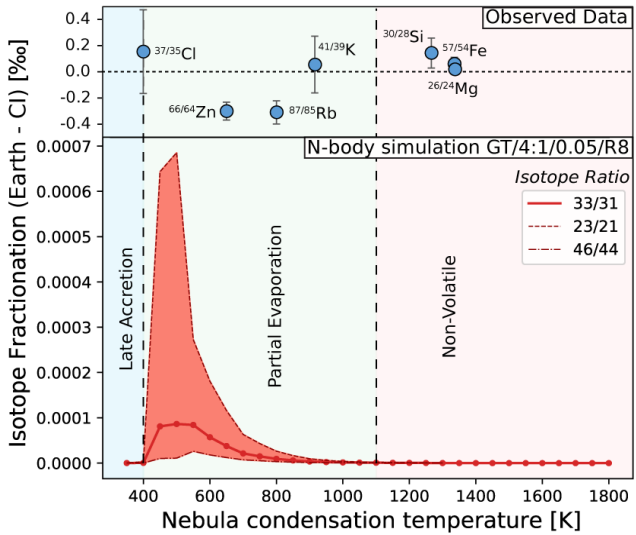

**Figure 4 | Stable isotope fractionation modelled in the fully accreted Earth as a function of volatility relative to observations.** Isotopic fractionation for elements for which the difference between Earth and CI chondrites is well-constrained, Cl[10], Zn[8], Rb[11], K[9], Si[42], Fe[43] and Mg[44], is shown in the top panel. The model results are displayed in the bottom panel, where the red line indicates the mean isotopic fractionation (in per mille relative to the initial value, *i.e.*, CI chondrites) across 1000 Monte Carlo runs. The isotope ratios in g/mol ($m_i$ = numerator isotope, $\Delta m_i = 2$ atomic mass units, amu) are listed in the legend. The vertical dashed lines delimit coloured zones pertaining to Late Accretion (blue, volatile elements whose isotopes are present in chondritic ratios), Partial Evaporation (green, elements that bear an isotopic signature of partial evaporation during atmospheric loss) and Non-Volatile (red, elements that were not lost and remain chondritic). The scale of isotopic variation observed exceeds that produced in the Monte-Carlo runs, except for low $m_i$ runs in N-body simulation GT/8:1/0.025/R2 (Extended Data Fig. 10). Moreover, the observations do not follow the behaviour expected for partial evaporation, which most readily generates heavy isotope ratios for elements with $450 < T_c^{50}$ (K) $< 800$.

Such isotope fractionation patterns are explicable through the dynamics of planetary impacts. The tendency for atmospheric loss (and hence isotope fractionation) to occur is quantified by the escape parameter, $\lambda_{esc}$ (eq. 5). As the Earth grows, $\lambda_{esc}$ gradually increases, upon which are superimposed embryo collisions ($M_E > 0.05$) that produce transient decreases in $\lambda_{esc}$ *via* impact heating (Extended Data Fig. 5), during which atmospheric loss is most pronounced. For any given N-body simulation, there are typically one- to two collisions that induce hydrodynamic escape. These events are restricted to the nascent stages of Earth's formation, during which its mass is ~0.05 to 0.2 $M_E$ (Extended Data Fig. 6). Even though temperatures in later, giant impacts may be higher, the proclivity for atmospheric loss is offset by the larger escape velocity, as $\lambda_{esc}$ depends on $v_{esc}^2$, while it depends on $1/T$ (eq. 5).

As such, chemical and isotopic fractionation generated by early, high-energy events are readily overprinted by continued accretion thereafter. The most volatile elements are delivered late by material that has seen no evaporation, meaning their isotopes retain chondritic ratios (Fig. 4). The isotope composition of elements with $450 < T_c^{50}$ (K) $< 800$ depends largely on the Earth's accretion path and hence on the N-body simulation. A small fraction (<5 %) of runs with the lowest $m_i$ in the most energetic simulation, GT/8:1/0.025/R2, reach 3 ‰ (Extended Data Fig. 10; 13). For all other runs, the limited degree of isotopic fractionation is consistent with the observed <0.4%[8–11] deviation of the BSE from CI chondrites in K (~1000 K), Rb (~800 K), Zn (~700 K) and Cl (~400 K) isotope compositions (Fig. 4). Therefore, volatile loss by partial atmospheric escape from large bodies (≥0.2 $M_E$) had minimal influence on the final compositional and isotopic make-up of Earth. By extension, isotopic variations of less volatile elements, including Mg, Fe and Si[42–44] (Fig. 4) are unlikely to originate from partial evaporation during collisions between embryos. Instead, they could have been inherited from an earlier stage, either during nebular condensation[42] or planetesimal evaporation[45]. Nevertheless, models that consider the specific thermodynamic properties of these elements over a wider range of N-body simulations are required. Conversely, the isotopic imprint of atmospheric loss is more likely to be preserved on small bodies for which escape is both more pronounced and less overprinted by continued accretion, consistent with the enrichment in heavy isotopes of K, Cl and Zn compositions of the Moon and Vesta[46–48].

Aluminium-26, which decays to $^{26}$Mg ($t_{1/2}$ = 0.7 Myr), was abundant enough to have caused melting and vaporisation on small rocky bodies, provided they formed sufficiently early (≤2 Myr)[49]. Therefore, a variation on the stochastic accretion hypothesis arises if the temperatures of precursor bodies reflect $^{26}$Al decay, either in addition to, or, in the extreme case, instead of the smooth heliocentric decrease in disk temperature as modelled above. The endmember case is investigated by randomly assigning each body an initial temperature selected from a normal distribution with mean, $T_0$ = 1000(±50, 1σ) K and standard deviation, σ = 175 K (Extended Data Fig. 5b). The $T_0$ value can reach ~1600 K, consistent with numerical models[50], and replaces $\dot{M}$ as the first random variable in the MC simulations. The remainder of the simulation is identical to that described above.

This temperature distribution also explains the abundance pattern of the BSE (Extended Data Fig. 14). However, because the compositions of planetesimals are independent of heliocentric distance, there is no systematic distinction in the accretion times of volatile and non-volatile elements, a result that contravenes evidence from the Pd-Ag and Hf-W systems[34]. This inconsistency can be resolved by numerical models that predict slower accretion of planetesimals with increasing heliocentric distance, causing the effects of $^{26}$Al heating to diminish[38,49]. In this scenario, a planetesimal population in the inner disk is predicted to have formed at 0.6 – 1.3 Myr[38]. Therefore, although differential $^{26}$Al heating perturbs the thermal structure imposed by the disk, it also produces the chemical and isotope composition of Earth by stochastic accretion, provided a heliocentric gradient is present.

The formation of Earth *via* pebble accretion[51], a popular alternative mechanism for planetary growth[52], leads to fundamentally different results. Here, pebble accretion is simulated by prescribing a mass accretion rate that scales with $M_E^{2/3}$, typical of 2D-pebble accretion in the Hill regime[53]. Pebble fluxes are assumed to mirror mass accretion rates to the Sun (eq. 3) and result in the growth of Earth to 0.9 $M_E$ in 3.85 Myr, consistent with disk accretion models (~5 Myr)[5]. Pebbles initially form at the snow line[4] and subsequently drift inwards by interactions with the surrounding nebular gas. Because pebbles are small, they thermally equilibrate with the gas at 1 AU, whose temperature is calculated over time with eqs. (2, 3). Thus, elemental abundances in pebbles at the time of their accretion by the Earth are determined as per planetesimals, but using the temperature at 1 AU as $T_{midplane}$.

Because pebbles accrete continuously in time, Earth's surface reaches only modest temperatures compared to planetesimal accretion, ~1750 K, sufficient to melt and vaporise the pebbles. However, the escape of vaporised material occurs in the upper atmosphere, near the Hill radius, or deeper, at the Bondi radius, if Earth is still embedded in the disk[54]. At these locations, temperatures drop to ~400- to 100 K[5], meaning elements with $T_c^{50}$ >400 K are solids. Consequently, atmospheric loss and isotopic fractionation is muted (Extended Data Fig. 15). Pebble accretion models produce concave-down patterns that decline at $T_c^{50}$ lower than observed in the BSE (Extended Data Fig. 16a). This results from the dependence of accretion rate on $M_E^{2/3}$, meaning the majority (85 %) of Earth's mass accretes between 1 and 3.85 Myr, when temperatures at 1 AU are ≤500 K (Extended Data Fig. 16b). As such, depletion of elements with $T_c^{50}$ above this temperature is limited. Therefore, unless pebbles themselves are already impoverished in volatile elements, pebble accretion alone does not produce the volatile depletion pattern of the Earth.

The distribution of initial temperatures, and hence compositions of accreted planetesimals and embryos, controlled the composition of the Earth. In order to achieve sufficient temperatures in these precursor bodies, either by heating in the disk and/or $^{26}$Al decay, they must have formed by ~1 Myr, a timescale that is consistent with radiometric ages for volatile depletion[39,41] and protoplanetary disk models[37,38]. Whether precursor body compositions were set by nebular condensation or planetesimal evaporation could be discriminated by holistic models of the growth of planets from the dust- to embryo stages. In the context of the Grand Tack, and other models in which the Earth accretes from a radial distribution of material (Extended Data Fig. 4), some form of heliocentric temperature gradient is necessary to explain the late accretion of volatile-rich bodies (Fig. 2) that satisfy the near-chondritic radiogenic[34] and stable[8–11] (Fig. 4) isotope composition of volatile elements in the Earth.

The mean volatile content of a planet, quantified in its $T_0$, is expected to increase with heliocentric distance. Should a heliocentric gradient have been established by disk accretion, then a continuous decrease in $T_0$ of the terrestrial planets would be observed, whereas $^{26}$Al heating may produce a dichotomy in the compositions of inner- and outer Solar System planetesimals due to their differential formation times. As they grow, planets accrete material from an expanding range of heliocentric distances[2,3]. This translates into a greater distribution of $T_0$ in precursor bodies, implying that larger planets should have more gradual depletion patterns (high-σ) than smaller ones (low-σ) (Fig. 1). Moreover, smaller planets are more likely to preserve isotopic imprints of atmospheric loss, resulting in heavy stable isotope compositions of volatile elements relative to their building blocks. Potassium isotopes in the Earth, Mars, the Moon and Vesta bear out this expectation, as they become increasingly non-chondritic with decreasing planetary mass[55]. The compositions of the terrestrial planets, together with their masses and heliocentric distances, offer the means by which to test these predictions.

## Methods

### N-body simulations

N-body simulations proceed by perfect-merger impacts that result in a net increase in mass of larger bodies at the expense of smaller ones (oligarchic growth) to produce three- to four terrestrial planets with masses and semi-major axes akin to those observed at the present-day[24].

We treat four simulations, labelled GT/4:1/0.05/R8, GT/4:1/0.025/R7, GT/8:1/0.08/R18 and GT/8:1/0.025/R2. See Table 1 for the masses and distribution of embryos and planetesimals, and refs. 24,25 for a detailed description of the simulations.

### Disk structure and initial composition

All bodies (embryos and planetesimals) are assigned an average density of 3000 kg/m$^3$. They are prescribed to initially contain some fraction of the CI-normalised abundances ($f_i$) of 30 fictive elements ($i$) defined with $T_c^{50}$ (at $10^{-4}$ bar total pressure) ranging from 350 to 1800 K at 50 K intervals, where:

$$f_i = x_{i,body}/x_{i,initial}, \quad (1)$$

and $x$ corresponds to the mole fraction. Where $x_{i,body} = x_{i,initial}$, the body has its full complement of element $i$, and is equivalent to a 'CI-chondritic' composition. The initial temperature of the planetesimal or embryo is determined by its heliocentric distance. The temperature profile of a high-opacity nebular midplane in steady-state mass accretion is[26,27]:

$$T_{midplane} = \left(\frac{9\tau G M_\odot \dot{M}}{64\pi\sigma_B d^3}\right)^{\frac{1}{4}}, \quad (2)$$

where $\tau$ is the optical depth, and is given by $\tau = \kappa\Sigma/2$ (ref. 56), where $\kappa$ is the Rosseland mean opacity (~0.3 m$^2$/kg in the range 100 – 1500 K[57]) and the surface density of the gas, $\Sigma = 3500(\dot{M}/10^{-8})(d/\mathrm{AU})^{-3/4}$ in kg/m$^2$, suitable for the optically-thick accretion phase of the star, $\dot{M} > 10^{-9}$ solar mass/yr, in the planet-forming region[27,58]), $M_\odot$ the mass of the Sun, $\dot{M}$ the mass accretion rate, $\sigma_B$ the Stefan-Boltzmann constant, and $d$ the heliocentric distance of the body given in the N-body simulation. $\dot{M}$ is the <u>first random variable</u> for the Monte-Carlo simulations (see *Monte-Carlo simulations*).

We impose the constraint that, for planetesimals, if $T_{midplane} > T_c^{50}$ of element $i$, $f_i = 0$, otherwise, $f_i = 1$, defining a 'step-function' pattern. As embryos are larger bodies, they themselves are expected to have been comprised of planetesimals with step-functions. Hence, abundances in the embryos reflect the summation of a number, N = (M$_{embryo}$/M$_{planetesimals}$) of planetesimals, each with a $T_{midplane}$ sampled normally about the $T_{midplane}$ of the embryo, with standard deviation σ$_{embryo}$. If the σ$_{embryo}$ is low, the depletion pattern in the embryo is closer to a step-function, while high σ$_{embryo}$ leads to a smoother curve. σ$_{embryo}$ is the <u>second random variable</u> for the Monte-Carlo simulations (see *Monte-Carlo simulations*).

### Energy budget during impact heating and cooling

Collisions result in a temperature increase according to the conversion of kinetic energy to heat. The energy released upon impact of two bodies of masses $M_A$ and $M_B$ travelling at velocities $v_A$ and $v_B$, respectively, is the sum of their kinetic energies:

$$E_k = \frac{1}{2}M_A v_A^2 + \frac{1}{2}M_B v_B^2. \quad (3)$$

However, only a fraction of the total kinetic energy is converted into heat, depending on the characteristics of the impacting material and obliquity and mass of the impact[59,60], while the heated region depends on the amount of energy transmitted beyond the isobaric core of the body. These poorly-known terms are consolidated into a single empirical constant, $h$, which ≈0.5[61]. The temperature increase is the total heat distributed throughout the affected mass, divided by the heat capacity of the material $C_P$:

$$\Delta T = \frac{h}{C_P}\left(\frac{E_k}{M_{planet}} - \Delta H_{fus} - \Delta H_{vap}\right) = \frac{h}{C_P}\left(\frac{v_{imp}^2}{2} - \Delta H_{fus} - \Delta H_{vap}\right). \quad (4)$$

Where $C_P$ is that for molten peridotite near its liquidus (1800 J/kgK[62]), $M_{planet}$ is the final mass of the planet, and $v_{imp}$ is the impact velocity given in the N-body simulation. Temperature increases if the energy delivered is in excess of the latent heat of fusion ($\Delta H_{fus} = 4 \times 10^5$ J/kg at 1400 K[63]), then vaporisation ($\Delta H_{vap} = 5 \times 10^6$ J/kg at 2000 K[61]) and finally an upper limit at 6000 K, whereupon behaviour becomes supercritical[64,65]. The mass of material that experiences peak temperatures (the isobaric core) scales with impactor radius[44,61]. However, because N-body simulations represent a swarm of planetesimal impacts, the heated area is likely to have been uniformly distributed in the body.

Following heating to maximum temperatures, the planetary body cools by a combination of black body radiation, latent heat of vaporisation and gravitational potential energy, with the energy balance given by[66]:

$$\frac{dT}{dt} = \frac{\left(\frac{dm}{dt}\left(\frac{GM_{planet}}{r} + \Delta H_{vap}\right) + 4\pi r^2 \sigma\epsilon T_s^4\right)}{M_{planet}C_P}. \quad (5)$$

The contribution of the evaporative cooling and escape term, $\frac{dm}{dt}\left(\frac{GM_{planet}}{r} + \Delta H_{vap}\right)$, is small and is neglected. Here, $r$ is the planetary radius (assuming a radiating atmosphere from the surface), $\sigma$ the Stefan-Boltzmann constant, and the surface temperature, $T_s$. The emissivity, $\epsilon$, takes into account that planets are not perfect black bodies, and is set to 0.05 to replicate cooling timescales for terrestrial planets with thick atmospheres[67–69].

Temperature is calculated at 10 K cooling intervals throughout the simulation. This temperature is then fed in as an input to calculate the composition of the atmosphere (*Vapour pressure*), its subsequent loss at each temperature step (*Atmospheric loss*), and thus to re-calculate a new bulk composition (*Compositional evolution*) in a loop.

### Vapour pressures

For silicate materials, vaporisation reactions can be generalised as:

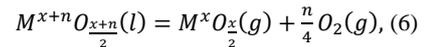

$$M^{x+n}O_{\frac{x+n}{2}}(l) = M^x O_{\frac{x}{2}}(g) + \frac{n}{4}O_2(g), \quad (6)$$

where $M^{x+n}O_{\frac{x+n}{2}}$ is the metal oxide species in the silicate melt, $M^x O_{\frac{x}{2}}$ is the stable gas species with metal oxidation state of $x$, balanced by $\frac{n}{4}$ moles of $O_2$ gas for a congruent reaction (*i.e.,* in which the gas(es) produced have the same composition as the reactant(s)).

At equilibrium, the partial pressure (assuming ideal gas behaviour such that $f = p$), is given by:

$$p\left(M^xO_{\frac{x}{2}}\right) = \frac{K_{(r)}X\left(M^{x+n}O_{\frac{x+n}{2}}\right)\gamma\left(M^{x+n}O_{\frac{x+n}{2}}\right)}{f(O_2)^{n/4}}, (7)$$

where,

$$K_{(r)} = \exp\left(\frac{-\Delta G_{(r)}}{RT_s}\right), (8)$$

for which $\Delta G_{(r)}$ is the Gibbs Free Energy of the vaporisation reaction, $R$ the gas constant and $T_s$ the equilibrium temperature set by the magma ocean at its surface in Kelvin. At low pressures in which volume changes are insignificant relative to the standard state (here taken at 1 bar),

$$\Delta G = \Delta H - T_s \Delta S \ (9)$$

In which $\Delta H$ is the enthalpy change and $\Delta S$ the entropy change of reaction. Substituting this equation into eq. (7), one obtains:

$$\log p(M^xO_{\frac{x}{2}}) = \left(\frac{\frac{-\Delta H}{T_s}+\Delta S}{2.303R}\right) + \log\left(\frac{X\left(M^{x+n}O_{\frac{x+n}{2}}\right)\gamma\left(M^{x+n}O_{\frac{x+n}{2}}\right)}{f(O_2)^{n/4}}\right). (10)$$

From experimental assessment of the equilibrium vapour pressures of metal-bearing gas species above metal oxides or metal oxide components in silicate melts, $\Delta S$ is near-constant[15,70], which results in slopes in log$p_i$ vs. $T$ space that are sub-parallel for almost all elements (*Extended Fig. 1*). Element volatilities are defined by their 50% condensation (or evaporation) temperature in the solar nebular gas[15]:

$$T_c^f = \frac{-\Delta H}{\left(R\left(\frac{n}{4}\ln fO_2\right) + \ln P_T + \ln\frac{f_{vap}}{(1-f_{vap})}\right) - \Delta S}. (11)$$

Here, $f_{vap}$ = the fraction in the vapour (=0.5 for 50% condensation). The $fO_2$ is that of a cooling solar nebula[71], while the total pressure is $10^{-4}$ bar in keeping with convention[19,29]. The $T_c^{50}$ of a given fictive element is defined by holding $\Delta S$ constant and changing the value of $\Delta H$ in eq. (11). Their properties are shown in *Extended Data Table 1*.

The partial pressure is determined as a function of temperature, element activity, $fO_2$ of the atmosphere and the stoichiometry of the vaporisation reaction, $n$ (eq. 10). This links the $T_c^{50}$ of an element with its general volatility behaviour under planet-forming conditions. Because most elements evaporate according to $n = 2$ stoichiometries[15], including all five major cation-forming elements in the BSE composition (Fe, Mg, Si, Ca, and Al), we adopt $n = 2$ for all fictive elements (*Extended Data Table 1*). Hence, relative fractionation by volatility is insensitive to $fO_2$, which is not the case for the alkalis ($n = 1$), or for the Group VI metals (Cr, Mo, and W) for which $n < 0$[15]. However, this simplification is apt to establish the overall trend of element depletion, and not to account for anomalies in the abundances of certain elements. Moreover, it is independent of any eventual revisions to $T_c^{50}$ values.

The sum of the partial pressures dictate the total pressure at any given $T$ and $fO_2$:

$$P_T = \Sigma_{M^xO_{\frac{x}{2}}=i}^i p\left(M^xO_{\frac{x}{2}}\right), (12)$$

since $p\left(M^xO_{\frac{x}{2}}\right)$ depends on $(fO_2)^{-n/4}$, congruent evaporation (eq. 6) demands that oxygen fugacity be equal to $n/4$ times the sum of the partial pressures of the metal oxide species. Because $n = 2$ in our treatment (*Extended Table 1*), then:

$$fO_2 = 0.5P_T = 0.5\Sigma_{M^xO_{\frac{x}{2}}=i}^i p\left(M^xO_{\frac{x}{2}}\right). (13)$$

Thus, the value of $fO_2$ is solved iteratively until the term $0.5P_T$ is equal to ½ the sum of the individual partial pressures of each component at a given temperature. Here, we constrain element activities by fitting our value of $P_T$ to that calculated for evaporation of silicate mantles[72]. To do so, we assign an activity to each fictive element as a function of its $T_c^{50}$ value. As per planetary materials, the three most abundant elements (Fe, Mg and Si) have $T_c^{50} \sim 1300 – 1400$ K. We use a Gaussian distribution to model a peak element activity about 1350 K. The rate at which element activity declines is found by minimisation to the $P_T$ determined in ref. 72 over the range $1800 < T$ (K) $< 3000$, using the objective function:

$$\chi^2 = \Sigma(P_T(ref. 72) - P_T(calc))^2. (14)$$

By iterating the value of $c$:

$$(x_i\gamma_i)_{initial} = \frac{\exp\left(-\frac{(T_{c,i}^{50}-1350)^2}{2c^2}\right)}{\Sigma_i(X_i\gamma_i)_{initial}}. (15)$$

We obtain a best-fit value to the total pressure above the bulk silicate Earth[72] (*Extended Data Fig. 2*) at $c = 100$, yielding the initial activities, $(x\gamma)_{initial}$, shown in *Extended Data Table 1*.

**Atmospheric loss**

Escape of the atmosphere is calculated self-consistently based on its pressure-temperature structure. The total pressure at the surface, $P_T$, (eq. 12) is a direct result of the planetary surface temperature, $T_s$, given that any crustal boundary layer is likely to be ~ a few cm thick[44,73] and prone to foundering. Even for small bodies ½ the mass of Pluto, the evaporation rate quickly attains a steady state with loss rate, such that the surface pressure is that at equilibrium[45]. A rapidly convecting terrestrial magma ocean (viscosities $\approx 0.1$ Pa.s[74]) experiences turnover in days, meaning evaporation is not diffusion-limited by transport of material to the surface.

Throughout Earth's accretion at any given temperature step, we model two thermal escape regimes, Jeans and hydrodynamic escape. Energy-limited or ionisation-driven escape are not considered as they require knowledge of solar energy flux. The prevailing mode of escape is determined by the competing influence of gravity and thermal energy, encapsulated in the escape parameter, $\lambda_{esc}$[75]:

$$\lambda_{esc} = \frac{v_{esc,exo}^2 m_i}{(2k_BT_{exo})}, (16)$$

where $v_{esc,exo}$ is the escape velocity of the planet at its exobase, $m_i$ is the average molar mass of the atmosphere, $k_B$ is the Boltzmann constant, and $T_{exo}$ is the temperature at the exobase. The quantity $m_i$ is the <u>third random variable</u> for the Monte-Carlo simulations (see *Monte-Carlo simulations*). Atmospheric structure is described by its scale height, $H$, the height over which its number density declines by a factor of $1/e$. For a well-mixed atmosphere, $H$ is evaluated at the tropopause by:

$$H_{trop} = \frac{k_BT_{trop}}{m_ig_s}\left(\frac{r_{trop}}{r_s}\right)^2, (17)$$

where,

$$g_s = \frac{GM_{planet}}{r_s^2}, (18)$$

where $r_s$ is the radius at the surface, and $r_{trop}$ is the radius at the tropopause ($r_{trop} = r_s + z_{trop}$), $M_{planet}$ is the mass of the planet. Thus, the temperature, $T_s$ and pressure, $P_T$ set by evaporation at the surface need to be extrapolated upwards to the tropopause. This is achieved using the expression for a convective, (dry adiabat) optically-thin troposphere[32]:

$$T_{trop} = T_s \left(\frac{P_{T,trop}}{P_T}\right)^{\frac{R}{C_P}}. (19)$$

Here, $R$ = the gas constant and $C_P$ is the heat capacity of an ideal, complex gas at constant pressure ($=4R$). For an optically-thin atmosphere in the absence of external heat sources, the temperature tends to a constant value given by the 'skin temperature'[32] ($=T_{trop}$),

$$T_s = 2^{0.25} T_{trop}. (20)$$

Combining eq. 20 with eq. 19 leads to $P_{T,trop} = 0.5\,P_T$. This limit, the tropopause, defines the transition from the convective troposphere to the diffusive stratosphere that overlies it. The adiabatic expansion of an ideal gas in hydrostatic equilibrium as a function of height above the surface ($z$) is:

$$P_T(z) = P_T \exp\left(-\frac{z}{H_z}\right), (21)$$

where $z$ at the tropopause is $z_{trop}$, and is therefore given by:

$$z_{trop} = -\ln(0.5) H_{trop}. (22)$$

As $z_{trop}$ appears in both eq. 22 and eq. 17, these equations are resolved iteratively until they converge upon a unique value of $z_{trop}$ and $H_{trop}$. Above this point, in the stratosphere, temperature as a function of height ($z$) is constant, reflecting the 'skin temperature'. The number density at the exobase, which is defined as the point at which the mean free path ($\ell = 1/n_{exo}\sigma$) equals the scale height ($H_{trop}$), is:

$$n_{exo} = \frac{1}{(H_{trop}\sqrt{2}\pi\sigma^2)}, (23)$$

where $\sigma$ is the collision integral:

$$\sigma = 1.5 \times 10^{-10}\,m. (24)$$

The escape velocity at the exobase is then calculated:

$$v_{esc,exo} = \sqrt{\frac{2GM_{planet}}{r_{exo}}} (25)$$

Where the radius of the exobase is given by adiabatic expansion:

$$r_{exo} = r_s + z_{exo} = r_s - \ln\left(\frac{n_{exo}}{n_s}\right) H_{trop} (26)$$

The number density at the surface is calculated by the ideal gas law:

$$n_s = \frac{P_T}{T_s k_B} (27)$$

The mean molar mass of the atmosphere, $m_i$, is held constant throughout the simulation. The escape parameter, $\lambda_{esc}$, at the exobase of the body can then be evaluated. If $\lambda_{esc} > 3$, the position of the exobase is further out than the radius at which the gas exceeds the sound speed, and Jeans escape prevails. In this regime, the mass loss rate is:

$$\frac{dn_i}{dt}\frac{1}{(4\pi r_{exo,i}^2)} = 0.73 \frac{n_{exo,i}}{2\sqrt{\pi}} \left(\frac{2k_B T_{eff}}{m_i}\right)^{0.5} (1 + \lambda_{esc}) \exp(-\lambda_{esc})$$
$$(molecules/m^2.s) (28)$$

or, in kg/s ;

$$\frac{dm_i}{dt} = 2.92 \pi r_{exo,i}^2 \frac{n_{exo,i}}{2\sqrt{\pi}} \left(\frac{2k_B T_{eff}}{m_i}\right)^{0.5} (1 + \lambda_{esc}) \exp(-\lambda_{esc}) m_i$$
$$(29)$$

If $\lambda_{esc} < 3$, the gas exceeds the speed of sound and a transonic outflow of the atmosphere results. The gas velocity is determined by the temperature ($T$), total pressure ($P_T$) and density ($\rho$) of its constituents at any given point, constrained by mass and momentum conservation[76]. At the 'sonic point', the velocity of the gas reaches that of the speed of sound in the medium ($C_s$). For an ideal gas, $C_s$ is given by:

$$C_s = \sqrt{\frac{\frac{7}{5}P_{T,trop}}{\rho_{trop}}} = \sqrt{\frac{\frac{7}{5}k_B T_{trop}}{m_i}}. (30)$$

Where the factor 7/5 is the value of the adiabatic index for a diatomic gas (5 degrees of freedom), and accounts for the heat of compression. The sonic point, or Bondi radius is:

$$r_B = \frac{GM_{planet}}{2C_s^2}. (31)$$

The hydrodynamic mass loss rate depends on the density of the gas phase at the Bondi radius. In order to estimate this value, the pressure of the gas at the surface is integrated outwards to $r_B$. For an adiabatically expanding gas, the relation is:

$$P_{r_B} = P_T \exp(-(r_B - r_{planet})/H_{r_B}) (32)$$

The number density (units of atoms/m$^3$) at the sonic point is:

$$n_{r_B,i} = P_{r_B}/k_B T_{r_B} (33)$$

The molecular loss rate (molecules/s) is then:

$$\frac{dn_i}{dt} = 4\pi r_B^2 n_{r_B,i} C_s (34)$$

or, in terms of kg/s ;

$$\frac{dm_i}{dt} = 4\pi r_B^2 n_{r_B,i} C_s m_i (35)$$

In order to estimate isotopic fractionation during atmospheric loss, two simulations are run concurrently by varying the term $m_i$ (which is a bulk property of the atmosphere in the model framework and cannot be independently varied for each element) according to:

$$m_j = m_i + (i - j) \times 1.67 \times 10^{-27} kg (36)$$

Here, the quantity $(i - j)$ is set to 2, reflecting the isotopic mass difference between most common systems (*e.g.*, $^{66}Zn/^{64}Zn$; $^{30}Si/^{28}Si$, $^{41}K/^{39}K$). The scale height of both isotopes, $i$ and $j$ are calculated independently, such that eq. 17 becomes:

$$H_{trop} = \frac{k_B T_{trop}}{m_j g_s} \left(\frac{r_{trop}}{r_s}\right)^2 (37)$$

simulating the separation of two isotopes by their scale heights in the stratosphere. Therefore, both Hydrodynamic and Jeans escape may cause isotope fractionation. Jeans escape rate is further modified due to the dependence of the relative mass loss rate $dm_i/dm_j$ on $(m_j/m_i)^{0.5}$. As such, eq. 28 becomes:

$$\frac{dn_j}{dt}\frac{1}{(4\pi r_{exo}^2)} = 0.73 \frac{n_{j,exo}}{2\sqrt{\pi}} \left(\frac{2k_B T_{eff}}{m_j}\right)^{0.5} (1 + \lambda_{esc}) \exp(-\lambda_{esc}) \quad (38)$$

The mean molar mass factor used to convert the loss rate from molecules/s to kg/s in both Jeans and Hydrodynamic escape is left constant ($m_i$), as $m_j$ is assumed to be an infinitely dilute component of the atmosphere only (i.e., $x_i \gg x_j$). Thus, the theoretical isotope fractionation is calculated for each element, presuming they all have the same value of $m_i$ and $m_j$.

**Compositional evolution**

The activity, $a$, of each element, $i$, in any planetary body is given by:

$$a_i = \frac{\left(\frac{(x_i \gamma_i)_{planet}}{(x_i \gamma_i)_{initial}}\right)}{\sum_i a_{i_{planet}}}, \quad (39)$$

where $(x_i \gamma_i)_{initial}$ is given for each $i$ by a Gaussian distribution centred at 1350 K, whose coefficients are adjusted such that the total pressure evolved as a function of temperature fits that for thermodynamic models for vaporisation of a bulk silicate Earth composition[72] (*Table A1*, eq. 15). For simplicity, here we assume ideal solution of each element, $\gamma_i = 1$. In the entire planet, therefore;

$$\sum_i a_i = \sum_i x_i = 1 \quad (40)$$

The composition and total pressure of the gas phase in equilibrium with the planet at each temperature step is calculated according to the section *Vapour pressure calculations*, a proportion of which is lost by Jeans or hydrodynamic escape, for which a mass flux is determined (*i.e.*, a loss of mass over a time interval, see *Atmospheric loss*). As the mass of the planet is known, this can be expressed in terms of the total planetary mass, which gives the normalised loss rate in seconds:

$$\frac{d\hat{m}}{dt} = \frac{\left(\frac{dm}{dt}\right)_{loss}}{M_{planet}} \quad (41)$$

In the model, the temperature is evaluated for a fixed time interval, $dt$, dictated by the time taken for the planet to cool by 10 K. Hence, multiplying the normalised loss rate by the seconds elapsed during the cooling interval gives the integrated mass loss (expressed as fraction of planetary mass) over that time interval:

$$d\hat{m} = \frac{d\hat{m}}{dt} dt \quad (42)$$

The new activity of a given element is:

$$a_{i+1} = a_i - d\hat{m}\frac{p_i}{P_T}, \quad (43)$$

Where $\frac{p_i}{P_T}$ is the mole fraction of the element $i$ in the gas phase being lost. Here, we assume that there is no fractionation of elemental mole fractions in the gas at the loss surface with respect to that set by equilibrium with the magma ocean at the planetary surface. That is, we neglect processes in the stratosphere that may fractionate elements according to their scale heights because the molar mass of the atmosphere is assigned as a bulk property, not a sum across each fictive element.

The output composition is then normalised to the new total mass of the body *i.e.*,

$$\hat{a}_{i+1} = \frac{a_{i+1}}{(\hat{m} - d\hat{m})} \quad (44)$$

This process is repeated, manifest as a compositional evolution of the planet dictated by the element's volatility and the rate of atmospheric loss. For isotope fractionation, this process is performed for elements $i$ and $j$ simultaneously throughout, such that the isotope composition (in per mille) of the body can be calculated at any time according to:

$$^{j/i}\delta_{planet}(\permil) = \left(\frac{\hat{a}_j}{\hat{a}_i} - 1\right) \times 1000 \quad (45)$$

**Monte-Carlo simulations**

For each simulation, there are three random variables:

i) $\dot{M}$. The stellar mass accretion rate. It sets the initial abundances of all planetesimals and embryos. $\dot{M}$ is randomly varied according to a normal distribution with a mean of $1 \times 10^{-7} M_\odot$ and a standard deviation ($\sigma$) of $0.25 \times 10^{-7} M_\odot$ in accord with observations of T-Tauri stars[77]. These accretion rates yield maximum temperatures (at 0.7 AU) of $1380^{+145}_{-220}$ K (1 $\sigma$), bracketing the $T_c^{50}$ of the major rock-forming elements. This range of temperatures is a robust constraint because refractory lithophile elements ($T_c^{50} \geq 1400$ K) are unfractionated from one another in planetary materials[78], providing empirical evidence that there was no significant accretion of components with fractionated abundances of elements with $T_c^{50} \geq 1400$ K.

ii) $\sigma_{embryo}$. The standard deviation for the temperature distribution of planetesimals that comprise a given embryo. The standard deviation, $\sigma_{embryo}$, is randomly varied with a normal distribution with mean 87 K and $\sigma = 24$ K about the mean $T_{midplane}$ value of the embryo in each simulation. This physically relates to embryo accretion from planetesimals sourced from variably wide heliocentric feeding zones[31,79]. Low values of $\sigma_{embryo}$ result in a 'step-function' distribution of elements, similar to planetesimals, whereas high values of $\sigma_{embryo}$ pertain to accretion of planetesimals from a wider temperature range, leading to a smoother element distribution pattern in embryos as a function of $T_c^{50}$.

iii) $m_i$. The mean atmospheric molar mass. A bulk property, it is sampled randomly from a normal distribution with mean 33.5 g/mol and $\sigma = 3.5$ g/mol, with minima and maxima corresponding to the masses of Na(g) (23 g/mol) and SiO(g) / $CO_2$(g) (both 44 g/mol), respectively. These are likely to represent the limits of the most abundant elements in a high-temperature vapour in equilibrium with a silicate mantle[72].

Monte Carlo simulations are performed 1000 times for each N-body simulation. The three variables are fixed for each simulation, selected according to the range and standard deviation for each normal distribution. The entire process described above has been coded together in a *Python* script, and was run on the Ludwig-Maximilians-Universität (LMU) supercomputing cluster. Each simulation takes roughly 1 minute to complete (1000 simulations = 1000 minutes).

**Pebble Accretion Simulation**

To simulate the accretion of the Earth predominantly by pebbles[80], a simple model is presented, in which the Earth grows to 90 % of its current mass solely by pebble accretion in 3.85 Myr. This process imparts heating due to the release of gravitational potential energy; from which a surface temperature is determined.

Because pebbles are small (~cm-sized) they thermally equilibrate with the surrounding nebular gas at the locus of accretion, that is, at 1 AU for Earth. In order to calculate the temperature at 1 AU, an estimation is required of how the mass accretion rate to the Sun decays as a function of time. To do so, eq. 4 of ref. [28], based on observations of the mass accretion rates of T-Tauri stars, is used to calculate $\dot{M}$ (eq. 2). For each given time step ($t$), the corresponding value of $\dot{M}$ is inserted into eq. (2) to calculate the temperature of the accreting pebbles. Pebble fluxes are assumed to mirror mass accretion rates (eq. 2) and scale with $M_E^{2/3}$, typical of 2D-pebble accretion in the Hill regime[53]. Given a temperature, their composition is determined in a manner analogous to that for planetesimals (see *Disk structure and initial composition*). The first and second terms of eq. 2 are varied within their uncertainties in a Monte Carlo simulation to evaluate the sensitivity of Earth's composition to mass accretion rate.

Because eq. 2 predicts very high temperatures in the first few thousand years (up to 3800 K at 5000 yr), a saturation limit is placed at 1350 K, which affects only the first 1.5±0.5 % of material accreted to the Earth. The vapour pressures, atmospheric loss and compositional evolution of Earth is treated in an identical manner to that outlined above, determined by the surface temperature given in the pebble accretion simulation.

**Code Availability**

The code described in the Methods is available upon reasonable request from the corresponding author.

**Data Availability**

Raw data produced by the numerical models can be obtained from the corresponding author upon reasonable request. Interested parties may contact S.A.J. & A.M. to access the N-body- and pebble accretion simulations upon reasonable request. The authors declare that all other data supporting the findings of this study are available within the paper and its supplementary information files.

**Acknowledgements**

We thank three anonymous reviewers for encouraging consideration of alternative modes of setting precursor body temperatures that led to a more detailed discussion of isotope fractionation and accretion histories. Denton Ebel is thanked for comments on an earlier version of this work. We are particularly grateful to Luca Maltagliati for his efficient and thoughtful editorial handling of the manuscript. P.A.S. acknowledges support from the Swiss National Science Foundation (SNSF) *via* an Ambizione Fellowship (#180025), and appreciates discussions with R. Pierrehumbert, M. Schönbächler, T. Lichtenberg and K. Mezger. I.L.S. recognises support from DFG Fellowship STO1271/2-1. A. M. acknowledges support from the ERC advanced grant HolyEarth N. 101019380

**Author Contributions**

P.A.S. conceived the study, developed the governing equations, and assisted with writing the Python code. I.L.S. developed the Python code and ran simulations. S.A.J. performed the N-body simulations. A.M. performed the N-body and pebble accretion simulations. H.O.N. contributed to the conceptual development of the study. P.A.S. wrote the paper with input from A.M. and H.O.N. All authors contributed to the discussion of the results.

**Author Information**

Reprints and permissions information is available at www.nature.com/reprints. The authors declare no competing financial interests. Readers are welcome to comment on the online version of the paper. Publisher's note: Springer Nature remains neutral with regard to jurisdictional claims in published maps and institutional affiliations. Correspondence and requests for materials should be addressed to P.A.S. (paolo.sossi@erdw.ethz.ch).